\documentclass[conference]{IEEEtran}
\IEEEoverridecommandlockouts

\usepackage{multicol}
\usepackage{color}
\usepackage{cite}
\usepackage{balance}
\ifCLASSINFOpdf
\usepackage[pdftex]{graphicx}
\usepackage{epstopdf}
\usepackage{subfigure}
\usepackage{pgfplots,multicol}
\usepackage{amsmath,stmaryrd,pict2e,picture}
\else
\fi
\usepackage{amsmath,amssymb,amsthm}
\usepackage{mathtools}
\usepackage{pifont}
\newenvironment{sysmatrix}[1]
{\left(\begin{array}{@{}#1@{}}}
	{\end{array}\right)}

\newlength{\rowidth}
\AtBeginDocument{\setlength{\rowidth}{3em}}
\usepackage{mathrsfs}
\usepackage{algorithm, algorithmic}

\def\footnoterule{\relax%
	\kern-5pt
	\hbox to \columnwidth{\hfill\vrule width 0.75\columnwidth height 0.4pt\hfill}
	\kern4.6pt}
\makeatother
\IEEEoverridecommandlockouts

	\title{Sparse Array Transceiver Design for Enhanced Adaptive Beamforming in MIMO Radar}
	\author{\IEEEauthorblockN{Syed~A.~Hamza$^1$, Weitong Zhai$^2$, Xiangrong Wang$^2$ and
			Moeness~G.~Amin$^3$}
		\IEEEauthorblockA{$^1$School of Engineering, Widener University, Chester, PA 19013, USA\\
		$^2$School of Electronic and Information Engineering, Beihang University, Beijing, 100191, China\\
			$^3$Center for Advanced Communications, Villanova University, Villanova, PA 19085, USA\\
			Emails: \{shamza\}@widener.edu, \{wtzhai, xrwang\}@buaa.edu.cn, \{moeness.amin\}@villanova.edu}
\thanks{The work by W Zhai and X Wang is supported by National Natural Science Foundation of China under Grant No. 62071021 and No. 61827901.}}
\begin{document}
\maketitle
\begin{abstract}
Sparse array design aided by emerging fast sensor switching technologies can lower the overall system overhead by reducing the number of expensive transceiver chains. In this paper, we examine the active sparse array design enabling the maximum signal to interference plus noise ratio (MaxSINR) beamforming at the MIMO radar receiver. The proposed approach entails an entwined design, i.e., jointly selecting the optimum transmit and receive sensor locations  for accomplishing MaxSINR receive beamforming. Specifically, we consider a co-located multiple-input multiple-output (MIMO) radar platform with orthogonal transmitted waveforms, and examine antenna selections  at the transmit and receive arrays. The optimum active sparse array transceiver design problem is formulated as successive convex approximation (SCA) alongside the two-dimensional group sparsity promoting regularization. Several examples are provided to demonstrate the effectiveness of the proposed approach in utilizing the given transmit/receive array aperture and degrees of freedom for achieving MaxSINR beamforming.
\end{abstract}

\IEEEpeerreviewmaketitle
\section{Introduction}
Non-uniform sparse arrays have emerged as an effective technology which can be deployed in various active and passive sensing modalities, such as acoustics and radio frequency (RF) applications, including GPS\cite{Amin2016}.  Fundamentally, sparse array design seeks optimum system performance, under both noise and interference, while being cognizant of the limitations on cost and aperture. Within the RF beamforming paradigm, maximizing the signal to interference plus noise ratio (SINR) amounts to concurrently optimizing the array configuration and beamforming weights. Equivalently, the problem can be cast as continuously selecting the best antenna locations to deliver MaxSINR under time-varying environment. In lieu of constantly moving  antennas to new positions, a more practical and feasible approach is to have a uniform full array and then switch among antennas with a fixed number of front-end chains, yielding the MaxSINR sparse beamformer for a given environment.


The above environment-dependent approach differs  from the environment-independent sparse array design that seeks to increasing the  spatial autocorrelation lags and maximizing the  contiguous portion of the coarray aperture for a limited number of sensors. The main task, therein, is to enable  DOA estimation involving  more sources than the physical sensors  \cite{1139138, 5456168, AMIN20171, 7012090, 8683315}.  For beamforming applications, the environment-independent  design criteria typically strives to achieve desirable beampattern characteristics  such as broad main lobe  and minimum sidelobe levels and  frequency invariant beampattern  for wideband design  \cite{81155, 23483091a3104a7f8bd7d213f8ada7cc, 1071, MURINO1997177,Wangwang, 9257073}.   

Environment-dependent sparse receive beamformer  striving to achieve MaxSINR can potentially provide sparse configurations that improve  target detection and estimation accuracy  \cite{663844798989898, 6774934, 7944303, 8645552, b014, mypaper, 8061016, 8682266, 8892512}. In this paper, we consider MaxSINR receive beamformer design for  MIMO radar operation. The  beamformer design, in this case, jointly selects the optimum transmit and receive sensor locations  for implementing an efficient receive beamformer. We assume that transmit sensors  emit orthogonal waveforms. The MIMO radar implementing sensor-based orthogonal transmit waveforms  does not benefit from   coherent transmit processing gain achieved when using directional beamforming. In this context, the proposed approach is fundamentally different from the parallel sparse array MIMO beamforming designs, which primarily rely on optimizing the transmit sensor locations and the corresponding correlation matrix of the transmit waveform sequence \cite{5680595, 837875911, 8378710, 10.1117/12.2557923, 9114655}. The proposed approaches, therein, maximize the transmitted signal power towards the perspective target locations while minimizing the cross-correlation of the target returns to achieve efficient receive beamforming characteristics. Moreover, existing design schemes essentially pursue an uncoupled  transmit/receive design  which is in contrast to the proposed approach  that incorporates  sparse receiver design in conjunction with transmit array optimization. 
 
%


The proposed transceiver beamforming design problem, in essence, translates to configuring the transmit/receive  array with the corresponding optimum receiver beamforming  weights. To this end, we pursue a data dependent  approach that jointly optimizes the beamforming sensor positions and weights. The optimization only requires the knowledge of the perspective desired source locations and does not assume any  knowledge of the interfering sources. We optimally select $M_{t}$ from $M$ transmitters and $N_{r}$ from $N$ receivers. Such selection is a binary optimization problem, and is NP-hard. In order to avoid extensive computations associated with enumeration of all possible array configurations, we apply convex relaxation. The design problem is posed as SCA with two-dimensional reweighted mixed $l_{1,2}$-norm penalization to jointly invoke sparsity in the transmit and receive dimensions. 
The rest of the paper is organized as follows:  In the next section, we  state the problem formulation for the MIMO  beamformer design in the case of a pre-specified array configuration. Section \ref{Optimum sparse array design} elaborates on the  sparse array design by  semidefinite relaxation to jointly find the optimum sparse transmit/receive array geometry. In the subsequent section,  simulations are presented to demonstrate the offerings of the proposed sparse array transmit design. The paper ends with concluding remarks.

\section{Problem Formulation}
\label{Problem Formulation}
We consider a MIMO radar with $M$ transmitters and $N$ receivers illuminating the scene through omni-directional beampattern. This is typically achieved through  emissions of orthogonal  waveforms across the transmit array. Specifically, we pursue  a co-located MIMO arrangement with the transmit and receive arrays in close vicinity. As such, a far-field source is presented by equal angles of departure and arrival. Consider $K$ target sources arriving from $\{\theta_{s,k}\}$. The  received baseband data $\mathbf{x}(n) \in \mathbb{C}^{MN \times 1}$  after matched filtering at the $N$ element uniformly spaced receiver array at  time instant $n$  is given by,
{\setlength\abovedisplayskip{5pt}
\setlength\belowdisplayskip{5pt}
\begin {equation} \label{a}
\mathbf{x}(n)=   \sum_{k=1}^K s_k(n) \mathbf{b}(\theta_{s,k}) +  \sum_{l=1}^{L_c} c_l(n)\mathbf{b}(\theta_{j,l})  + \sum_{q=1}^{Q}  \mathbf{i}_{q}(n)  + \mathbf{v}(n),
\end {equation}}where, $s_k(n) \in \mathbb{C}$  is the $k$th reflected target signal. The extended steering vector $\mathbf{b}(\theta)$ of the virtual array is $\mathbf{b}(\theta)=\mathbf{a}_t(\theta) \otimes \mathbf{a}_r(\theta)$. In the case of uniform transmit and receive linear arrays with a respective inter-element spacing of $d_t$ and $d_r$, the transmit and receive steering vectors are given by, 
{\setlength\abovedisplayskip{5pt}
\setlength\belowdisplayskip{5pt}
  \begin{equation}
  \label{b}
 \mathbf{a}_t(\theta)= {}[1 \,  \,  \, e^{j 2 \pi (d_t/\lambda) cos\theta}  {}\\
 \,  . \,   . \,  . \, \,  \,  \, \,  \,  \, e^{j 2 \pi (M-1) (d_t/\lambda) cos\theta}]^T,
 \end{equation}}
 and
 {\setlength\abovedisplayskip{5pt}
\setlength\belowdisplayskip{5pt}
   \begin{equation}
  \label{b}
 \mathbf{a}_r(\theta)= {}[1 \,  \,  \, e^{j 2 \pi (d_r/\lambda) cos\theta}  {}\\
 \,  . \,   . \,  . \, \,  \,  \, \,  \,  \, e^{j 2 \pi (N-1) (d_r/\lambda) cos\theta}]^T.
 \end{equation}}
The variance of additive Gaussian noise $\mathbf{v}(n)\in \mathbb{C}^{MN \times 1}$ is  $\sigma_{v}^2$ at the receiver output. There are $L_c$ interferences $c_l(n)$ mimicking target reflected signal and $Q$  narrowband interferences   $\mathbf{i}_q (n)$ $\in \mathbb{C}^{MN \times 1}$. The latter  is defined as the Kronecker product of the receiver steering vector $\mathbf{a}_r(\theta_{i,q})$ and the matched filtering output of the interference $\mathbf{j}_q(n)$ such that $\mathbf{i}_q(n)=\mathbf{j}_q(n) \otimes \mathbf{a}_r(\theta_{i,q})$.
 The received data vector $\mathbf{x}(n)$  is then linearly combined to maximize the output SINR. The output signal $y(n)$ of the optimum beamformer for MaxSINR is given by \cite{1223538}, 
 {\setlength\abovedisplayskip{5pt}
\setlength\belowdisplayskip{5pt}
 \begin {equation}  
 \label{c}
 y(n) = \mathbf{w}^H \mathbf{x}(n)
 \end {equation}}where $\mathbf{w}$ is the beamformer weight at the receiver. The optimal solution $\mathbf{w}_{o}$ can be obtained by solving the optimization problem that seeks to minimize the interference power at the receiver output while preserving the desired signal. The constraint minimization problem can be cast as,
 {\setlength\abovedisplayskip{5pt}
\setlength\belowdisplayskip{5pt}
 \begin{equation} \label{d}
 \begin{aligned}
 \underset{\mathbf{w} \in \mathbb{C}^{MN}}{\text{minimize}} & \quad   \mathbf{w}^H\mathbf{R_{x}}\mathbf{w},\\
 \text{s.t.} & \quad     \mathbf{ w}^H\mathbf{R}_s \mathbf{ w} = 1,  
 \end{aligned}
 \end{equation}}where the source correlation matrix is $\mathbf{R}_{s}$ $=$ $\sum_{k=1}^K\sigma_{s,k}^2$ $\mathbf{b}(\theta_{s,k})\mathbf{b}^H(\theta_{s,k})$, with $\sigma^2_{s,k} =E\{s_k(n)s_k^H(n)\}$ denoting the average received power from the $k$th target return. The data correlation matrix, $\mathbf{R}_{\mathbf{x}}\approx ( 1/T)\mathbf{x}(n)\mathbf{x}(n)^H$, is directly estimated   from the $T$ received data snapshots. The  solution to the   optimum weights  in (5)  is given by $\mathbf{w}_{o} =\mathscr{P} \{  \mathbf{R}_{\mathbf{x}}^{-1} \mathbf{R}_{s} \}$, with the operator $\mathscr{P} \{. \}$  representing the principal eigenvector of the input matrix. This optimum solution yields the MaxSINR, SINR$_{o}$, given by \cite{1223538},
{\setlength\abovedisplayskip{5pt}
\setlength\belowdisplayskip{5pt}
 \begin{equation}  \label{f}
 \text{SINR}_{o} = \Lambda_{max}\{\mathbf{R}^{-1}_{n} \mathbf{R}_{s}\},
 \end{equation}}which is the MaxSINR is the maximum eigenvalue ($\Lambda_{max}$) of  the product of  the two matrices, the inverse of interference plus noise correlation matrix  and the desired source correlation matrix. It is clear that the resulting solution holds irrespective of the array configuration, whether the array is uniform or sparse. For the latter, the performance of MaxSINR beamformer is intrinsically tied to the array configuration.  The sparse optimization of the above formulation is explained in the next section. 
 
\section{Sparse array design}  
\label{Optimum sparse array design} 
The separate sensor selection problem for a joint transmit and receive design is a combinatorial optimization problem and can't be solved in polynomial time.  We formulate the sparse array design problem by exploiting the structure of the received signal model and  solve it by applying sequential convex approximation. To exploit the sparse structure of the joint transmit and receive sensor selection, we introduce a  two dimensional $l_{1,2}$-mixed norm regularization to recover group sparse solutions. One dimension pertains to the transmitter sparsity, whereas the other sparsity pattern is associated with the receiver side. Moreover, this two-dimensional sparsity pattern is entwined and coupled, meaning that when one transmitter is discarded, all $N$ receiving data pertaining to this transmitter is zero. Similarly, the receiver is discarded only when its received data corresponding to all the $M$ transmitters is zero. The structure of the optimal sparse  beamforming weight  vector is elucidated in (7), where $\checkmark$ denotes a sensor location activated, or selected, and $\times$ denotes a sensor not activated.
{\setlength\abovedisplayskip{5pt}
\setlength\belowdisplayskip{5pt}
\begin{alignat}{2}
\overbrace{\underbrace{\begin{sysmatrix}{rrrr}
		\checkmark \\
		\times  \\
		\times  \\
		\checkmark \\
		\vdots \\
		\times  
		\end{sysmatrix}}_{\text{(Tx 1 active)}  }}^{ \mathbf{w}(1,N)} \quad
\overbrace{\underbrace{\begin{sysmatrix}{rrrr}
		\checkmark \\
		\times  \\
		\times  \\
		\checkmark \\
		\vdots \\
		\times  
		\end{sysmatrix}}_{(\text{Tx 2 active)}  }}^{ \mathbf{w}(N+1,2N)} \quad
\overbrace{\underbrace{\begin{sysmatrix}{rrrr}
		\times \\
		\times  \\
		\times  \\
		\times \\
		\vdots \\
		\times  
		\end{sysmatrix}}_{(\text{Tx 3 inactive)}  }}^{ \mathbf{w}(2N+1,3N)} \quad \hdots
\overbrace{\underbrace{\begin{sysmatrix}{rrrr}
		\checkmark \\
		\times  \\
		\times  \\
		\checkmark \\
		\vdots \\
		\times  
		\end{sysmatrix}}_{(\text{Tx $M$ active)}  } }^{ \mathbf{w}(N(M-1)+1,MN)}
\end{alignat}}In (7), each column vector denotes the receive beamformer weights for a fixed transmit location. It is noted that the optimal sparse beamformer, corresponding to the active  transmit and receive locations, follows a group sparse structure along the transmit and receive dimensions (the consecutive $N$ sensors are discarded vertically or the consecutive $M$ sensors are discarded horizontally. It is evident that the missing transmit sensor at position $3$, for example, translates to the sparsity along all the  corresponding  $N$ entries of  $\mathbf{w}$ ($N$ consecutive $\times$ vertically in (7)). Similarly, the group sparsity is also invoked across the received signals corresponding to all transmitters ($M$ consecutive $\times$ horizontally in (7)). 

\subsection{Group Sparse solutions through SCA}
The problem in (5) can equivalently be rewritten by swapping the objective and constraint functions as follows,
{\setlength\abovedisplayskip{5pt}
\setlength\belowdisplayskip{5pt}
\begin{equation} \label{i2}
\begin{aligned}
\underset{\mathbf{w \in \mathbb{C}}^{MN}}{\text{maximize}} & \quad   \mathbf{ w}^H\bar{\mathbf{R}}_s\mathbf{ w}\\
\text{s.t.} & \quad     \mathbf{w}^H\mathbf{R_x}\mathbf{ w} \leq 1 
\end{aligned}
\end{equation}}where $\bar{\mathbf{R}}_s=-\mathbf{R}_s$. The sparse MIMO configuration of uniformly spaced receivers and transmitters with a respective inter-element spacing of $d_r$ and $d_t=Nd_r$ is employed for data collection. The covariance matrix $\mathbf{R_x}$ of the full receive virtual array can be then obtained by the matrix completion method proposed in \cite{b014}. 

The beamforming weight vectors are generally complex valued, whereas the  quadratic functions are real. This observation  allows expressing the problem with  only real variables which is typically accomplished by  replacing the correlation matrix $\bar{\mathbf{R}}_s$ by $\tilde{\mathbf{R}}_s$  and concatenating the beamforming weight vector accordingly \cite{Ibrahim2018MirrorProxSA}, 
{\setlength\abovedisplayskip{5pt}
\setlength\belowdisplayskip{2pt}
\begin{multline} \label{k2}
\tilde{\mathbf{R}}_s=\begin{bmatrix}
\text{real}(\bar{\mathbf{R}}_s)       & -\text{imag}(\bar{\mathbf{R}}_s)   \\
\text{imag}(\bar{\mathbf{R}}_s)       & \text{real}(\bar{\mathbf{R}}_s)   \\
\end{bmatrix},
\tilde{\mathbf{w}}=\begin{bmatrix}
\text{real}({\mathbf{w}})       \\
\text{imag}({\mathbf{w}})   \\
\end{bmatrix}
\end{multline}}
Similarly, the received data correlation matrix $\mathbf{R_x}$ is replaced by $\mathbf{\tilde{R}_x}$. The problem in  (\ref{i2}) then  becomes,
{\setlength\abovedisplayskip{5pt}
\setlength\belowdisplayskip{2pt}
\begin{equation} \label{l2}
\begin{aligned}
\underset{\mathbf{\tilde{w} \in \mathbb{R}}^{2MN}}{\text{minimize}} & \quad   \mathbf{\tilde{ w}}^{'}\tilde{\mathbf{R}}_s\mathbf{\tilde{ w}},\\
\text{s.t.} & \quad     \mathbf{\tilde{ w}}^{'}\mathbf{\tilde{R}_x}\mathbf{\tilde{ w}} \leq 1,
\end{aligned}
\end{equation}}where $'$ denotes transpose operation. After expressing the constraint in terms of real variables, we convexify the objective function  by utilizing  the first order approximation iteratively,
{\setlength\abovedisplayskip{5pt}
\setlength\belowdisplayskip{2pt}
\begin{equation} \label{m2}
\begin{aligned}
\underset{\mathbf{\tilde{ w} \in \mathbb{R}}^{2MN}}{\text{minimize}} & \quad   \mathbf{m}^{(k)}{'}\mathbf{\tilde{ w}}+b^{(k)},\\
\text{s.t.} & \quad     \mathbf{\tilde{ w}}^{'}\mathbf{\tilde{R}_x}\mathbf{\tilde{ w}} \leq 1,
\end{aligned}
\end{equation}}where, $\mathbf{m}^{(k)}$ and ${b^{(k)}}$, updated at the $k+1$ iteration, are given by $\mathbf{m}^{(k+1)}=2\tilde{\mathbf{R}}_s\mathbf{\tilde{ w}}^{(k)}, b^{(k+1)}=-\mathbf{\tilde{w}}^{(k)}{'}\tilde{\mathbf{R}}_s \mathbf{\tilde{w}}^{(k)} $, respectively. Finally, to invoke sparsity in the beamforming weight vector, the re-weighted mixed $l_{1,2}$ norm is adopted primarily for promoting group sparsity,  
{\setlength\abovedisplayskip{0pt}
\setlength\belowdisplayskip{-5pt}
\begin{align} \label{n2}
\underset{\mathbf{\tilde{w}}, \mathbf{c},\mathbf{r}}{\text{minimize}} \quad & \quad   \mathbf{m^{(k)}{'}}\mathbf{\tilde{ w}}+b^{(k)}  +   \alpha(\mathbf{p}^{'}\mathbf{c})  +\beta(\mathbf{q}^{'}\mathbf{r}) \\
\text{s.t.} 
& \quad \mathbf{\tilde{ w}}^{'}\mathbf{\tilde{R}_x}\mathbf{\tilde{ w}}\leq 1, \tag{12a}\\
& \quad ||\mathbf{P}_{i}\odot  \mathbf{\tilde{w}}||_{2}\leq c_{i}, \tag{12b}\\
& \quad 0\leq c_{i}\leq 1,\quad i=1,...,M \tag{12c}\\
& \quad ||\mathbf{Q}_{j}\odot \mathbf{\tilde{w}}||_{2}\leq r_{j}, \tag{12d}\\
& \quad 0\leq r_{j}\leq 1,\quad j=1,...,N \tag{12e}\\
& \quad \mathbf{1}_{M}^{'}\mathbf{c}=M_{t}, \tag{12f}\\
& \quad \mathbf{1}_{N}^{'}\mathbf{r}=N_{r}, \tag{12g}
\end{align}}
and
{\setlength\abovedisplayskip{5pt}
\setlength\belowdisplayskip{-20pt}
\begin{align} \label{n2}
\mathbf{P}_{i}=[\hspace{-3mm}\overbrace{0  ...0...0}^{\mbox{\footnotesize$\begin{array}{c}N\ elements\\
1st\ group \end{array}$}}\hspace{-3mm} ...\hspace{-3mm} \overbrace{1...1...1}^{\mbox{\footnotesize$\begin{array}{c}N\ elements\\
 ith\ group \end{array}$}} \hspace{-3mm}...\hspace{-3mm} 
 \overbrace{1...1...1}^{\mbox{\footnotesize$\begin{array}{c}N\ elements\\
 (M+i)th\ group \end{array}$}} \hspace{-2mm}...\hspace{-3mm}\overbrace{0...0...0}^{\mbox{\footnotesize$\begin{array}{c}N\ elements\\
    2Mth\ group \end{array}$}}\hspace{-3mm}]^{'},
\end{align}}



{\setlength\abovedisplayskip{1pt}
\setlength\belowdisplayskip{5pt}
\begin{align} \label{n2}
\mathbf{Q}_{j}=[\ \overbrace{\underbrace{0\  ...\ 0}_{(j-1)\ 0s}\ 1\ 0\ ...\ 0}^{\mbox{\footnotesize$\begin{array}{c}
N\ elements\ of\\
the\ 1st\ group \end{array}$}}\ .\ .\ . \ \overbrace{\underbrace{0\  ...\ 0}_{(j-1)\ 0s}\ 1\ 0\ ...\ 0}^{\mbox{\footnotesize$\begin{array}{c}
N\ elements\ of\\
    the\ 2Mth\ group \end{array}$}}]^{'}.
\end{align}}

Here, $\odot$ means the element-wise product, $\mathbf{c}$ and $\mathbf{r}$ are two auxiliary binary selection vectors, $\mathbf{P}_{i} \in \{0,1\}^{2MN}$ in (12b) is the transmission selection matrix, which is used to select the real and imaginary parts of the $N$ weights corresponding to the $i$th transmitter, as shown in (13). Eqs. (12c) and (12f) indicate that we can select $M_{t}$ transmitters at most. Similarly, $\mathbf{Q}_{j}$ in (12d) is the receiver selection matrix, as shown in (14). Eqs. (12e) and (12g) indicate that we can select $N_{r}$ receivers at most. The two parameters $\alpha$ and $\beta$ are used to control the sparsity of the transmitters and the receivers, $\mathbf{p}$ and $\mathbf{q}$ are the reweighting coefficient vectors for the transmitters and receivers respectively and the detailed update method is given below.
 
\begin{algorithm}[t!] \label{algorithm}
\setlength{\abovecaptionskip}{5pt}
\setlength{\belowcaptionskip}{5pt}
	
	\caption{SCA-Sparse Transmit/Receive Beamformer Design}
	
	\begin{algorithmic}[]
		
		\renewcommand{\algorithmicrequire}{\textbf{Input:}}
		
		\renewcommand{\algorithmicensure}{\textbf{Output:}}
		
		\REQUIRE  $M$, $N$, $M_{t}$, $N_{r}$, $\alpha =\beta=0.5$, $\alpha_0 =\beta_0=1$, target direction $\theta_s$, auto-correlation matrix $\mathbf{R}_{s}$ and $\mathbf{R}_{x}$   \\
		
		\ENSURE  Optimal weight $\mathbf{w}$, location and number of receivers and transmitters.   \\
		
		\textbf{Initialization:} \\
		
		Initialize $\mathbf{w}$, $\mathbf{p}$, $\mathbf{q}$, $\mathbf{m}$ as all ones matrix. Set $\alpha_{0}$ = 1 and $\beta_{0}$ = 1. Initialize the tradeoff parameters $\alpha$ and $\beta$ according to the sparsity requirement.
		\WHILE {$||\mathbf{\tilde{w}}^{(k+1)}-\mathbf{\tilde{w}}^{(k)}||_{2}\geq 10^{-5}$}
		\STATE   a)  Convert $\mathbf{w}$, $\mathbf{R}_{s}$ and $\mathbf{R}_{x}$ to the real domain to get $\mathbf{\tilde{w}}$, $\mathbf{\tilde{R}}_{s}$ and $\mathbf{\tilde{R}}_{x}$ according to Eq.(9).\\
		\STATE   b)  Update $\mathbf{p}^{(k+1)}$ and $\mathbf{q}^{(k+1)}$ according to Eq.(13) \\
    	\STATE   c)  Update $\mathbf{\tilde{w}}^{(k+1)}$ according to Eq.(12). \\ 
		\STATE   d)  Convert $\mathbf{\tilde{w}}^{(k+1)}$ back into the complex solution $\mathbf{w}^{(k+1)}$ by $\mathbf{w}^{(k+1)}(i)=\mathbf{\tilde{w}}^{(k+1)}(i)+j\mathbf{\tilde{w}}^{(k+1)}(i+MN)$.\\
		
		\ENDWHILE
		
		\RETURN $\mathbf{w}$
		
	\end{algorithmic}
	\label{algorithm}
\end{algorithm}

\subsection{Reweighting Update}
\label{Reweighting Update}

A common method of updating the reweighting coefficient is to take the reciprocal of $|\mathbf{w}|$ \cite{Candes2008}. However, this can't control the number of elements to be selected. Thus, similar to \cite{WANG2020102684}, in order to make the number of selected antennas controllable, we update the weights using the following formula,
{\setlength\abovedisplayskip{5pt}
\setlength\belowdisplayskip{5pt}
\begin{equation} \label{g2}
{p}_i^{(k+1)}=\frac{1-c_i^{(k)}}{1-e^{-\beta_{0}c_i^{(k)}}+\epsilon}-(\frac{1}{\epsilon})(c_i^{(k)})^{\alpha_{0}}
\end{equation}}

Here, $\alpha_{0}$ and $\beta_{0}$ are two parameters that control the shape of the curve, and the parameter $\epsilon$  avoids the unwanted explosive case. The essence of this re-weight update method is to take a large positive penalty for a entry close to zero and a small negative reward for the entry close to 1. As a result, the entries of the two selection vectors $\mathbf{c}$ and $\mathbf{r}$ tend to be either 0 or 1. The update of the re-weight vector $\mathbf{q}$ follows the same rule. Through iterative regression, we can find $M_{t}$ transmitters and $N_{r}$ receivers. In addition, by controlling the values of $\alpha$ and $\beta$, we can obtain different sparsities of transmitters and receivers \cite{6477161}.  The proposed algorithm for joint transmit/receive beamformer design is elaborated further in  Algorithm 1.

 \section{Simulations} \label{Simulations}
In this section, we demonstrate the effectiveness of our proposed sparse MIMO radar from the perspective of output SINR. We compare the performance of the optimal MIMO array transceiver configured according to the proposed algorithm with randomly configured MIMO arrays. In practice, the interference may be caused by co-existence in the same bandwidth or being deliberately positioned at some angles transmitting the same waveform as targets of interest. We consider both kinds of interferences.  

\subsection{Example 1}
In this example, we fixed the direction of the interferences, and change the arrival angle of the target from $0^{\circ}$ to $90^{\circ}$. We consider a full uniform linear MIMO array consisting of 8 transmitters (M = 8) and 8 receivers (N = 8). We  select $M_t=4$ transmitters and $N_r=4$ receivers. For the receiver, we set the minimum spacing of the sensors to $\lambda/2$, while for the transmitter, we set the minimum spacing of the sensors to $N\lambda/2$. Suppose there are two interferences with an interference to noise-ratio (INR) of 13dB and arrival angles of $\theta_{q}=[40^{\circ},70^{\circ}]$. The SNR of the desired signal is fixed at 20dB. We simulate  co-existing interferences and deliberate interferences. The output SINR versus target angle is shown in Fig.1. In this figure, case 1 corresponds to two deliberate interferences, whereas case 2 represents two co-existing interferences. It can be seen that, in both cases, the optimal sparse MIMO arrays exhibit better performance than randomly selected sparse arrays. Fig.2, depicts  the configuration of the optimal MIMO array for a target at $80^{\circ}$.

\subsection{Example 2}
In this example, we consider the scenario where the interferences are spatially  close to the target which causes a great adverse impact on the array performance. We change the angle of the target from $0^{\circ}$ to $90^{\circ}$. For each target angle, two interferences are generated from the proximity of $\pm 5^{\circ}$ away from the target. We consider a full uniform linear MIMO array consisting of 20 transmitters (M = 20) and 20 receivers (N = 20). We select 5 transmitters and 5 receivers to compose the sparse MIMO array. The other simulation parameters remain the same as those in example 1. Again, we plot the output SINR versus target angle, as shown in Fig. 3. In case 1, the two interferences are deliberate, whereas in case 2, both interferences are co-existing. 
It can be observed that when the interferences are spatially close to the target, the proposed optimal MIMO sparse array exhibits more superiority than randomly configured sparse MIMO arrays.


\begin{figure}[!t]
	\setlength{\abovecaptionskip}{-0.1cm}
	\centering
	\includegraphics[height=2.8in, width=3.5in]{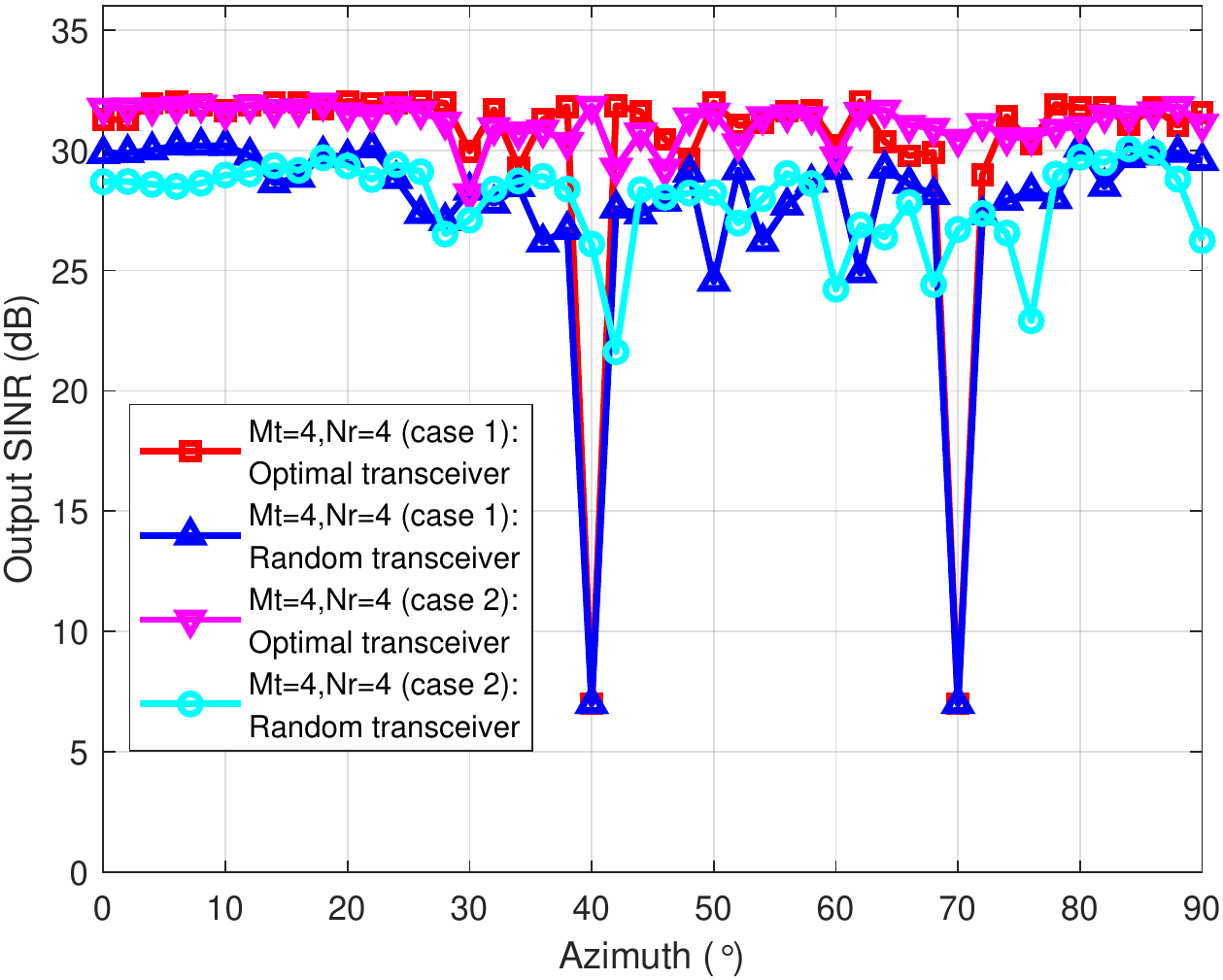}
	\caption{Relationship between output SINR and target angle with the directions of interferences being fixed.}
	\label{Fig.1}
\end{figure} 

\begin{figure}[!t]
	\setlength{\abovecaptionskip}{-0.2cm}
	\centering
	\includegraphics[height=2.4in, width=3.5in]{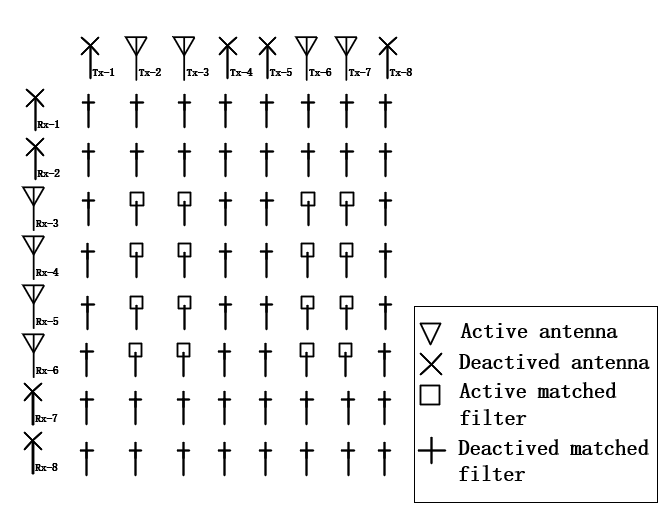}
	\caption{Optimal MIMO transceiver configuration when the target is from $80^{\circ}$.}
	\label{Fig.2}
\end{figure}

\begin{figure}[!t]
	\setlength{\abovecaptionskip}{-0.1cm}
	\centering
	\includegraphics[height=2.8in, width=3.5in]{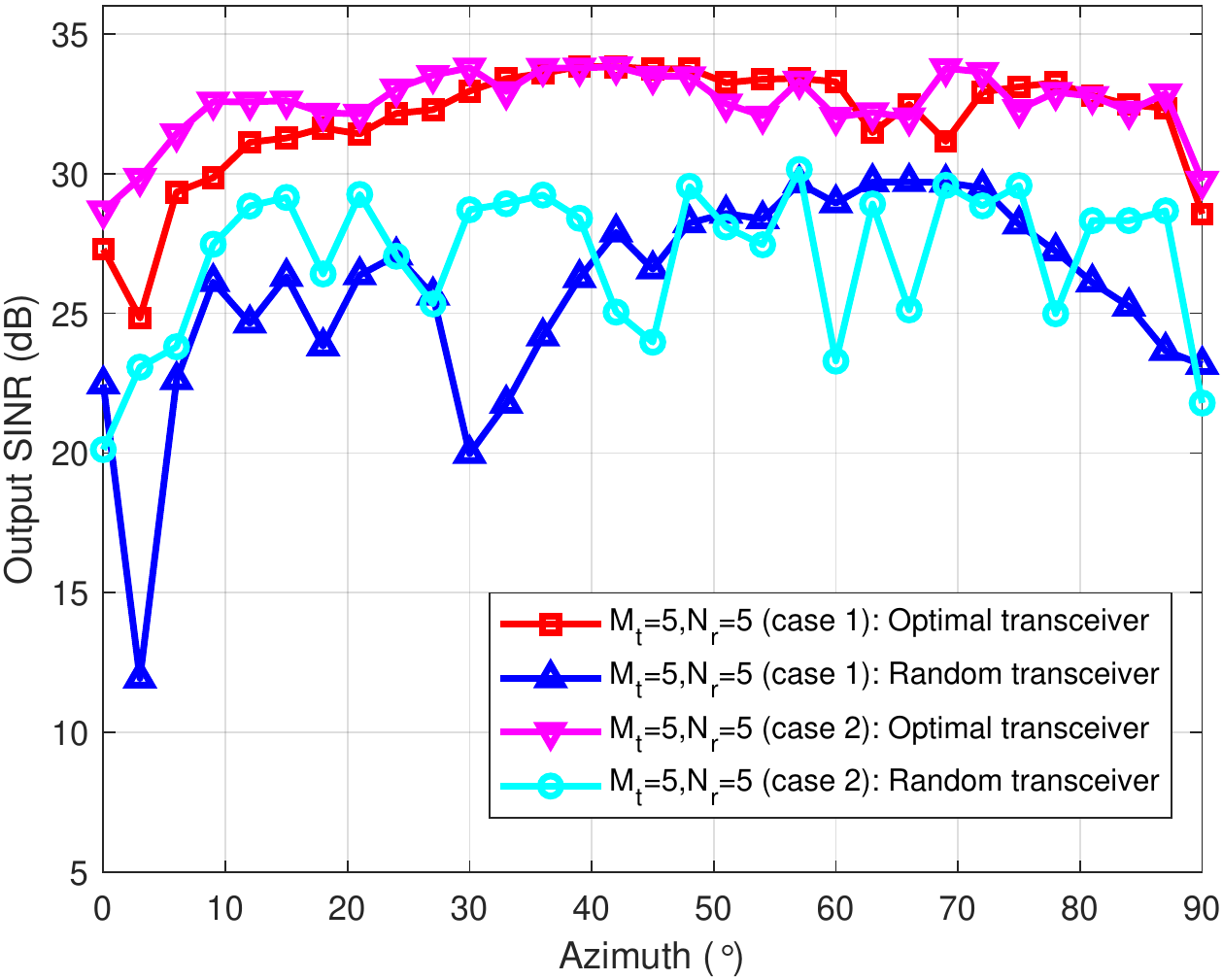}
	\caption{Relationship between the output SINR and the target angle when the interferences are closed to the target.}
	\label{Fig.3}
\end{figure}

\section{Conclusion}
In this paper, the problem of sparse transceiver design for MIMO radar was considered. For a given number of transmitting and receiving sensors, a sparse MIMO array structure in terms of MaxSINR was jointly designed.  A mixed-norm reweighted regularization was utilized to promote two-dimensional sparsity. It was shown by simulation examples that the proposed sparse MIMO transceiver selection algorithm  provides superior interference suppression performance to randomly designed array. 


\begin{figure*}
\begin{multicols}{2}
\bibliographystyle{IEEEtran}
\bibliography{references}
\end{multicols}
\end{figure*}

\end{document}